\documentclass[12pt]{article}
\usepackage{epsfig}
\usepackage{amsmath}
\usepackage{hhline}
\usepackage{amssymb}
\usepackage{times}
\usepackage{cite}

\newlength{\dinwidth}
\newlength{\dinmargin}
\begin{document}  
% The rest
\newcommand{\pom}{{I\!\!P}}
\newcommand{\reg}{{I\!\!R}}
\newcommand{\slowpi}{\pi_{\mathit{slow}}}
\newcommand{\fiidiii}{F_2^{D(3)}}
\newcommand{\fiidiiiarg}{\fiidiii\,(\beta,\,Q^2,\,x)}
\newcommand{\n}{1.19\pm 0.06 (stat.) \pm0.07 (syst.)}
\newcommand{\nz}{1.30\pm 0.08 (stat.)^{+0.08}_{-0.14} (syst.)}
\newcommand{\fiidiiiful}{F_2^{D(4)}\,(\beta,\,Q^2,\,x,\,t)}
\newcommand{\fiipom}{\tilde F_2^D}
\newcommand{\ALPHA}{1.10\pm0.03 (stat.) \pm0.04 (syst.)}
\newcommand{\ALPHAZ}{1.15\pm0.04 (stat.)^{+0.04}_{-0.07} (syst.)}
\newcommand{\fiipomarg}{\fiipom\,(\beta,\,Q^2)}
\newcommand{\pomflux}{f_{\pom / p}}
\newcommand{\nxpom}{1.19\pm 0.06 (stat.) \pm0.07 (syst.)}
\newcommand {\gapprox}
   {\raisebox{-0.7ex}{$\stackrel {\textstyle>}{\sim}$}}
\newcommand {\lapprox}
   {\raisebox{-0.7ex}{$\stackrel {\textstyle<}{\sim}$}}
\def\gsim{\,\lower.25ex\hbox{$\scriptstyle\sim$}\kern-1.30ex%
\raise 0.55ex\hbox{$\scriptstyle >$}\,}
\def\lsim{\,\lower.25ex\hbox{$\scriptstyle\sim$}\kern-1.30ex%
\raise 0.55ex\hbox{$\scriptstyle <$}\,}
\newcommand{\pomfluxarg}{f_{\pom / p}\,(x_\pom)}
\newcommand{\dsf}{\mbox{$F_2^{D(3)}$}}
\newcommand{\dsfva}{\mbox{$F_2^{D(3)}(\beta,Q^2,x_{I\!\!P})$}}
\newcommand{\dsfvb}{\mbox{$F_2^{D(3)}(\beta,Q^2,x)$}}
\newcommand{\dsfpom}{$F_2^{I\!\!P}$}
\newcommand{\gap}{\stackrel{>}{\sim}}
\newcommand{\lap}{\stackrel{<}{\sim}}
\newcommand{\fem}{$F_2^{em}$}
\newcommand{\tsnmp}{$\tilde{\sigma}_{NC}(e^{\mp})$}
\newcommand{\tsnm}{$\tilde{\sigma}_{NC}(e^-)$}
\newcommand{\tsnp}{$\tilde{\sigma}_{NC}(e^+)$}
\newcommand{\st}{$\star$}
\newcommand{\sst}{$\star \star$}
\newcommand{\ssst}{$\star \star \star$}
\newcommand{\sssst}{$\star \star \star \star$}
\newcommand{\tw}{\theta_W}
\newcommand{\sw}{\sin{\theta_W}}
\newcommand{\cw}{\cos{\theta_W}}
\newcommand{\sww}{\sin^2{\theta_W}}
\newcommand{\cww}{\cos^2{\theta_W}}
\newcommand{\trm}{m_{\perp}}
\newcommand{\trp}{p_{\perp}}
\newcommand{\trmm}{m_{\perp}^2}
\newcommand{\trpp}{p_{\perp}^2}
\newcommand{\alp}{\alpha_s}

\newcommand{\alps}{\alpha_s}
\newcommand{\sqrts}{$\sqrt{s}$}
\newcommand{\LO}{$O(\alpha_s^0)$}
\newcommand{\Oa}{$O(\alpha_s)$}
\newcommand{\Oaa}{$O(\alpha_s^2)$}
\newcommand{\PT}{p_{\perp}}
\newcommand{\JPSI}{J/\psi}
\newcommand{\sh}{\hat{s}}
\newcommand{\uh}{\hat{u}}
\newcommand{\MP}{m_{J/\psi}}
\newcommand{\PO}{I\!\!P}
\newcommand{\xbj}{x}
\newcommand{\xpom}{x_{\PO}}
\newcommand{\ttbs}{\char'134}
\newcommand{\xpomlo}{3\times10^{-4}}  
\newcommand{\xpomup}{0.05}  
\newcommand{\dgr}{^\circ}
\newcommand{\pbarnt}{\,\mbox{{\rm pb$^{-1}$}}}
\newcommand{\gev}{\,\mbox{GeV}}
\newcommand{\WBoson}{\mbox{$W$}}
\newcommand{\fbarn}{\,\mbox{{\rm fb}}}
\newcommand{\fbarnt}{\,\mbox{{\rm fb$^{-1}$}}}
%
% Some useful tex commands
%
\newcommand{\qsq}{\ensuremath{Q^2} }
\newcommand{\gevsq}{\ensuremath{\mathrm{GeV}^2} }
\newcommand{\et}{\ensuremath{E_t^*} }
\newcommand{\rap}{\ensuremath{\eta^*} }
\newcommand{\gp}{\ensuremath{\gamma^*}p }
\newcommand{\dsiget}{\ensuremath{{\rm d}\sigma_{ep}/{\rm d}E_t^*} }
\newcommand{\dsigrap}{\ensuremath{{\rm d}\sigma_{ep}/{\rm d}\eta^*} }
% Journal macro
\def\Journal#1#2#3#4{{#1} {\bf #2} (#3) #4}
\def\NCA{\em Nuovo Cimento}
\def\NIM{\em Nucl. Instrum. Methods}
\def\NIMA{{\em Nucl. Instrum. Methods} {\bf A}}
\def\NPB{{\em Nucl. Phys.}   {\bf B}}
\def\PLB{{\em Phys. Lett.}   {\bf B}}
\def\PRL{\em Phys. Rev. Lett.}
\def\PRD{{\em Phys. Rev.}    {\bf D}}
\def\ZPC{{\em Z. Phys.}      {\bf C}}
\def\EJC{{\em Eur. Phys. J.} {\bf C}}
\def\CPC{\em Comp. Phys. Commun.}

%%% My specific commands
\newcommand{\etm}{\ensuremath{E_t^{\mathrm{miss}}}}
\newcommand{\nnga}{$\nu^* \rightarrow \nu \gamma$~}
\newcommand{\newqq}{$\nu^* \rightarrow e W_{\hookrightarrow q \bar{q}} $~}
\newcommand{\nnzqq}{$\nu^* \rightarrow \nu Z_{\hookrightarrow q \bar{q}} $~}
\newcommand{\md}{M_{\mbox{jj}}}

\begin{titlepage}

\noindent
DESY 01-145  \hfill  ISSN 0418-9833 \\
September 2001

\vspace{3cm}

\begin{center}
\begin{Large}
  {\bf 
	Search for Excited Neutrinos at HERA}\\

\vspace{2cm}

H1 Collaboration

\end{Large}
\end{center}

\vspace{2cm}

\begin{abstract}
\noindent
We present a  search for excited neutrinos
using $e^- p$ data taken by the H1 experiment at HERA at a center-of-mass energy of  $ 318 $~GeV with an integrated luminosity of 15 pb$^{-1}$.
No evidence for excited neutrino production is found. Mass dependent exclusion limits are determined for the ratio of the coupling to the compositeness scale, $f/\Lambda$, independently of the relative couplings to the SU(2) and U(1) gauge bosons.
%values of the SU(2) and U(1) form factors $f_s$, $f$ and $f'$. 
These limits extend the excluded region to higher masses than has been possible in previous searches at other colliders.
\end{abstract}

\vspace{1.5cm}

\begin{center}
Submitted to Physics Letters B
\end{center}

\end{titlepage}

%
%          COPY THE AUTHOR AND INSTITUTE LISTS 
%       AT THE TIME OF THE T0-TALK INTO YOUR AREA
%
% from /h1/iww/ipublications/h1auts.tex 

\begin{flushleft}
  %-- H1AUTS Author list by names 
%-- Status: Wed Jun 20 09:24:27 MET DST 2001  Number of authors = 335 

C.~Adloff$^{33}$,              %WUPP-ST        01/96           Adloff              
V.~Andreev$^{24}$,             %LPI -PD        8/88            Andreev             
B.~Andrieu$^{27}$,             %ECPL-PD        8/88            Andrieu             
T.~Anthonis$^{4}$,             %ANTW-ST        11/99           Anthonis            
V.~Arkadov$^{35}$,             %ZEUT-LEFT      10/0            Arkadov             
A.~Astvatsatourov$^{35}$,      %ZEUT-ST        02/98           Astvatsatourov      
A.~Babaev$^{23}$,              %ITEP-PD        8/88            Babaev              
J.~B\"ahr$^{35}$,              %ZEUT-PD        8/88            Baehr               
P.~Baranov$^{24}$,             %LPI -PD        8/88            Baranovp            
E.~Barrelet$^{28}$
% ...

%-- H1 Institutes
\bigskip{\it
 $ ^{1}$ I. Physikalisches Institut der RWTH, Aachen, Germany$^{ a}$ \\
% ...
$ ^{4}$ Inter-University Institute for High Energies ULB-VUB, Brussels;
          Universitaire Instelling Antwerpen, Wilrijk; Belgium$^{ c}$ \\
% ...
$ ^{23}$ Institute for Theoretical and Experimental Physics,
          Moscow, Russia$^{ l}$ \\
 $ ^{24}$ Lebedev Physical Institute, Moscow, Russia$^{ e,h}$ \\
% ...
$ ^{27}$ LPNHE, Ecole Polytechnique, IN2P3-CNRS, Palaiseau, France \\
 $ ^{28}$ LPNHE, Universit\'{e}s Paris VI and VII, IN2P3-CNRS,
          Paris, France \\
% ...
$ ^{33}$ Fachbereich Physik, Bergische Universit\"at Gesamthochschule
          Wuppertal, Wuppertal, Germany \\
% ...
$ ^{35}$ DESY, Zeuthen, Germany \\
% ...
$ ^a$ Supported by the Bundesministerium f\"ur Bildung und
      Forschung, FRG,
      under contract numbers 05 H1 1GUA /1, 05 H1 1PAA /1, 05 H1 1PAB /9,
      05 H1 1PEA /6, 05 H1 1VHA /7 and 05 H1 1VHB /5  \\
% ...
$ ^c$ Supported by FNRS-NFWO, IISN-IIKW \\
% ...
$ ^e$ Supported by the Deutsche Forschungsgemeinschaft \\
% ...
$ ^h$ Supported by Russian Foundation for Basic Research
      grant no. 96-02-00019 \\
% ...
$ ^l$ Partially Supported by Russian Foundation
      for Basic Research, grant    no. 00-15-96584 \\
}

\end{flushleft}

\newpage

The discovery of excited states of quarks or leptons,
as predicted by compositeness models\cite{Harari:1984xy,Boudjema:1993em}, would supply convincing evidence for a new substructure of matter.
Electron-proton interactions at very high energies provide ideal conditions
to look for excited states of first generation fermions. In particular 
a magnetic type coupling of the electron
would allow for the production of single excited neutrinos ($\nu^*$) through t-channel $W$ boson exchange. The phenomenology of this process is described in~\cite{Hagiwara:1985wt,Baur:1990kv,Adloff:2000gv}.
In this paper we present a search for $\nu^*$ production
  followed by the electroweak decays 
  $\nu^* \rightarrow \nu \gamma$, $\nu^* \rightarrow e W$ or $\nu^* \rightarrow \nu Z$. 
  The analysis makes use of 15 pb$^{-1}$ of $e^- p$ data with an electron beam energy of 27.6 GeV and a proton beam energy of 920 GeV collected in  1998 and 1999 with the H1 experiment at HERA.  
Compared to previous H1 results from $e^-p$ collisions~\cite{Ahmed:1994yi} the analysis benefits from an increase in luminosity by a factor of 30 and by an increase of the center-of-mass energy from 300 GeV to 318 GeV. Furthermore, it also improves significantly on results derived from larger luminosity of $e^+p$ data at a center-of-mass energy of 300 GeV~\cite{Adloff:2000gv}, due to a much larger cross-section for $\nu^*$ production in $e^-p$ scattering as compared to the $e^+p$ case.
At a $\nu^*$ mass of 200 GeV the ratio of those cross-sections is of
the order of 100. 
Other searches for excited neutrinos have recently been presented by
ZEUS~\cite{Chekanov:2001xk} and by LEP
experiments~\cite{Abreu:1999jw,Abbiendi:2000sa,Acciarri:2001kb}. 

The production cross section and the decays of excited neutrinos 
can be calculated using an effective Lagrangian~\cite{Hagiwara:1985wt,Baur:1990kv}
which depends on a compositeness 
mass scale $\Lambda$ and on form factors (reduced here to parameters) $f_s$, $f$ and $f'$ allowing for the composite lepton to have arbitrary coupling strengths associated to the gauge groups SU(3), SU(2) and U(1).
The excited neutrino can decay into the electroweak gauge bosons via 
$\nu^* \rightarrow \nu \gamma$, $\nu^* \rightarrow e W $ and $\nu^* \rightarrow \nu Z $.
As shown in~\cite{Baur:1990kv}, the decay width of the $\nu^*$ is a function of $f$, $f'$ and $\Lambda$ and  can reach, for part of the accessible mass range, a few hundred GeV, much larger than the detector resolution (10 GeV).
For smaller decay widths (corresponding to masses below 200 GeV) the narrow width approximation (NWA) is applicable, in which the assumption is made that the production and decay of a particle factorize. In this range the COMPOS~\cite{Kohler:1991yu} generator is used for cross-section calculations. For masses beyond 200~GeV the full cross-section for $\nu^*$ production and decay is evaluated with COMPHEP~\cite{Pukhov:1999gg} using the Lagrangian given 
in ~\cite{Baur:1990kv}. In the overlap region the compatibility of COMPOS and COMPHEP has been verified.

The detector components of the H1 experiment~\cite{Abt:1997hi} most relevant for this analysis are shortly described in the following.
The interaction region is surrounded by a system of drift and proportional
chambers covering the polar angular range\footnote{The polar angle $\theta$ is measured with respect to the proton beam direction ($+z$).} $7^o < \theta < 176^o$. The tracking system is
placed inside a finely segmented liquid argon (LAr) calorimeter
covering the polar angular range 4$^o~<~\theta~<$~154$^o$~\cite{Andrieu:1993kh}.
Energy resolutions of
 $\sigma_E / E \simeq 12\% / \sqrt{E(GeV)} \oplus 1\%$ for
 electrons and $ \sigma_E / E \simeq 50\% / \sqrt{E(GeV)} \oplus 2\%$ for
 hadrons have been obtained in test beam measurements 
~\cite{Andrieu:1994yn,Andrieu}. The tracking system and calorimeters are surrounded by a
 superconducting solenoid and an iron yoke instrumented with streamer tubes.
Leakage of hadronic showers outside the calorimeter is measured by analogue 
charge sampling of the streamer tubes with a resolution~\cite{iron} of $\sigma_E / E \simeq 100\% /~\sqrt{E(GeV)}$.

For the decays of the heavy gauge bosons only the dominating hadronic modes are considered.
The selection of $\nu^*$ events is based on photon or electron identification, missing
transverse energy (\etm) measurement and the requirement for jets,
depending on the channel investigated. Electromagnetic clusters are
requested to have more than 95\% of their energy in the
electromagnetic part of the calorimeter and to be isolated from other
particles~\cite{Schoening}. 
 They are further differentiated into electron 
and photon candidates using the data of 
associated charged tracks.
Jets with a minimum transverse momentum of 5~GeV are reconstructed from the hadronic final state in the LAr calorimeter using a cone algorithm, adapted from the LUCELL scheme in the JETSET package~\cite{Sjostrand:1995iq}.

Background not related to $e^- p$ collisions is rejected by requiring
a primary interaction vertex reconstructed within 
$\pm 35$ cm around the nominal vertex value, by using topological filters 
and by requiring the event time to coincide with the time of the bunch crossing.
Standard Model (SM) backgrounds which could mimic the $\nu^*$ signatures
are Neutral Current Deep Inelastic Scattering (NC~DIS), Charged Current Deep Inelastic Scattering (CC~DIS) and photoproduction processes ($\gamma p$). 
The background expectation from NC~DIS and CC~DIS is calculated using
the event generator DJANGO~\cite{Schuler:1991yg} which includes first
order QED corrections based on HERACLES~\cite{Kwiatkowski:1992es} and
QCD radiation based on the Colour Dipole
Model~\cite{Andersson:1989gp}. Parton densities are taken from the
MRST parameterization~\cite{Martin:1998sq} which includes constraints
from DIS measurements at HERA up to a squared momentum transfer $Q^2$
=
5000~$\rm{GeV}^2$~\cite{Aid:1996au,Adloff:1997mf,Derrick:1996ef,Derrick:1996hn}. 
The hadronisation process is simulated in the Lund string
fragmentation scheme using JETSET~\cite{Sjostrand:1995iq}.  
Direct and resolved $\gamma p$ processes, including prompt photon production, are simulated with PYTHIA~\cite{Sjostrand:1994yb}. 
All Monte Carlo samples are subject to a full simulation of the H1 detector.

The \nnga channel is characterized by missing transverse energy and by an electromagnetic cluster in the calorimeter. The main SM background is expected from CC~DIS.
Events are selected with an identified photon of transverse momentum
($P_t$) greater than 16 GeV and total missing transverse energy \etm  greater than 16 GeV.
To reject NC~DIS background where the scattered electron (sometimes misinterpreted as a photon) is preferably scattered through small angles, photon candidates are accepted in the forward region of the detector only ($\theta < 1.8$~rad). For \etm $>30$~GeV electromagnetic clusters in the
very forward region ($\theta < 1 $~rad) are accepted even if they
are linked to a track.
In this particular region the conversion rate $\gamma \rightarrow ee$ and also the number of randomly assigned tracks is expected to be higher due to the high multiplicity of hadronic charged particles from jets. 
In order to be able to reconstruct the event vertex position from charged particles, the event is required to contain a jet. To further suppress background from events in which hadronic energy fluctuations of jets result in a measured missing transverse momentum, the missing transverse momentum vector of the event is required to have a component of more than 8~GeV perpendicular to the required jet.
To reduce the influence of photons coming from QED radiation along the quark line, the jet must be isolated from the photon in azimuth ($\Delta\varphi($jet,$\gamma) > 0.35$ rad). 
In total 2~events are found in this channel for an expected background
of $ 3.0 \pm 0.2$~(stat.) $ \pm 1.2 $~(syst.) events. The different sources of systematic errors are discussed below. The background is composed of 2.7~events from CC~DIS
and 0.3~events from NC~DIS with negligible contributions from $\gamma p$. 
The resulting selection efficiency ranges between 40\% and 65\%.

The \newqq channel is characterized by an electromagnetic cluster with an associated track and
two jets. The main SM background is NC DIS as photoproduction events
do not yield a significant rate of electrons with high transverse
momentum ($P_t^{ele}$). A cut $P_t^{ele} >$ 12.5~GeV is chosen.
At very high transverse momentum $P_t^{ele} >$ 85~GeV the background
from NC DIS is low and no further cuts are applied. In the range
65~GeV $< P_t^{ele} <$ 85~GeV two jets with an invariant mass $\md >$ 50~GeV are
required as expected for a hadronic $W$ decay. In the range 12.5~GeV $< P_t^{ele} <$ 65~GeV three jets are required, 
where the third jet is supposed to originate from the quark struck in the $W$ proton interaction. A cut on the electron polar angle is applied which depends on $P_t^{ele}$ and ranges from $\theta_e < 1.20 $ to $\theta_e < 2.25$~rad. To reconstruct a
$W$ candidate, the dijet-pair with invariant mass closest to the nominal $W$
boson mass is accepted in the range
65~GeV~$< \md <$~87~GeV. 
The two jets chosen as the $W$ candidate are ordered by their transverse
momentum such that $P_t^{jet1}>P_t^{jet2}$. As for many background events 
jet~2 points in the very forward direction, an additional cut on its polar angle, $\theta^{jet2}>$ 0.2~rad, is applied if the transverse momentum of this jet is lower than 30 GeV. 
After these cuts, 6 events remain in
the data. The expected background is $ 7.0 \pm 0.6 \pm 1.4 $ events mainly from NC DIS with negligible contributions from CC DIS and $\gamma p$. The resulting signal efficiency ranges between 30\% and 50\%.

The \nnzqq channel is characterized by two
jets and missing transverse energy  \etm . The main background is expected from 
CC~DIS with a moderate contribution from $ \gamma p $, whereas the NC~DIS
contribution is sufficiently suppressed for large  \etm . A cut
$ $~\etm~$ > 10$~GeV is chosen. At $ $~\etm~$ > 40$~GeV only two jets are
required, while at lower \etm a third jet is required and events with
an electron or photon candidate are rejected. A $Z$ candidate is reconstructed from the combination of 2 jets  with invariant mass closest to the 
nominal $Z$ boson mass provided this mass is greater than 76 GeV. Again 
these two jets are ordered in $P_t$. To suppress
further the background from CC~DIS a cut on
the  polar angle $\theta^{jet2}>$ 0.15 rad is applied. In the region of
relatively low missing transverse momentum $10$~GeV~$<$~\etm~$<$~20~GeV
an additional cut is applied on the transverse momentum of jet 1
($P_t^{jet1}>$50 GeV).
With these criteria, one candidate event is found in the
data, with an expected background of $ 3.7 \pm 0.2 \pm 0.9 $ events. The background consists
mainly of CC DIS (2.3 events) and $\gamma p$ (1.3 events). The
resulting signal efficiency is above 60\% for masses greater than 150 GeV.

Contributions to the systematic uncertainties come from the limited knowledge of
the absolute energy scale of the calorimeter and missing higher order corrections in the
event generators which are used for the background estimation. The
uncertainties of the electromagnetic energy scale amount to 0.7\% in the
central part of the detector and up to 3\% in the forward region. For the
hadronic part an uncertainty of 4\% is assigned. For the \nnga channel
the lack of QED radiation from the quark line in the DJANGO
generator leads to an uncertainty of the CC DIS background expectation which, 
after applying the $\Delta\varphi($jet,$\gamma) > 0.35$ rad cut, is limited to 
 $40\%$ as estimated using ~\cite{wwga}.
For the \newqq and \nnzqq channels the background normalization
is varied by 15\% to
account for differences observed in particular for the 3-jets production 
between perturbative calculations of the
order $O(\alpha_s^2)$~\cite{Carli:1998zr,bate,tampere157} and the parton shower approach. The statistical error of the Monte Carlo event samples is taken into account. Finally, the luminosity measurement leads to a normalization uncertainty of 2.25\%.

In all three search channels the number of observed and expected events are
in good agreement. Upper limits at 95$\%$ confidence level on the
coupling $f/\Lambda$ are thus derived  
as described in \cite{Adloff:2000gv}
following the Bayesian approach~\cite{Barnett:1996hr,Helene:1984ph}.
\begin{figure}[tb]
\begin{center}

\begin{tabular}{cc}
 
\epsfxsize=0.48\textwidth
 \epsffile{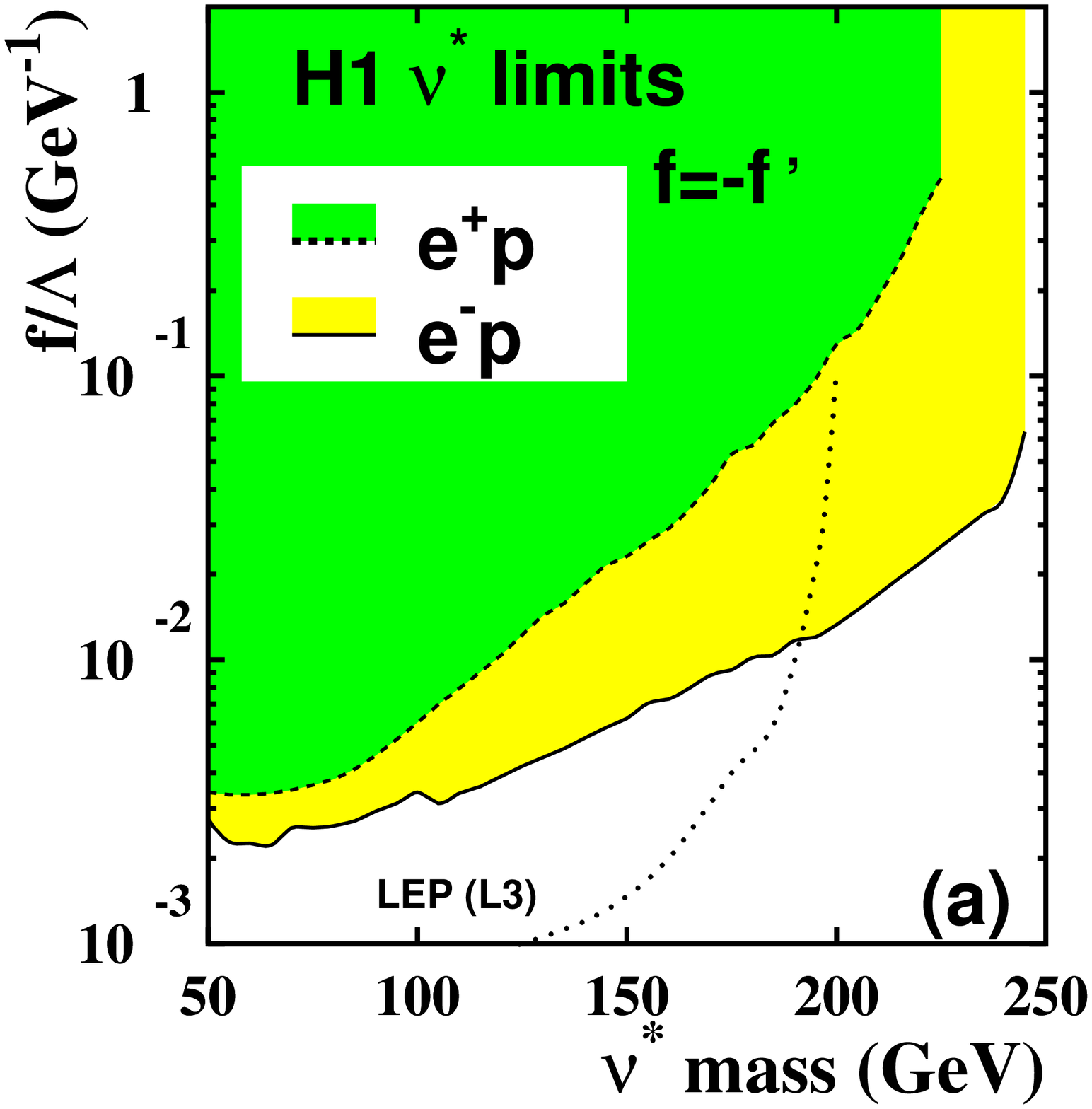} &
\epsfxsize=0.48\textwidth
\epsffile{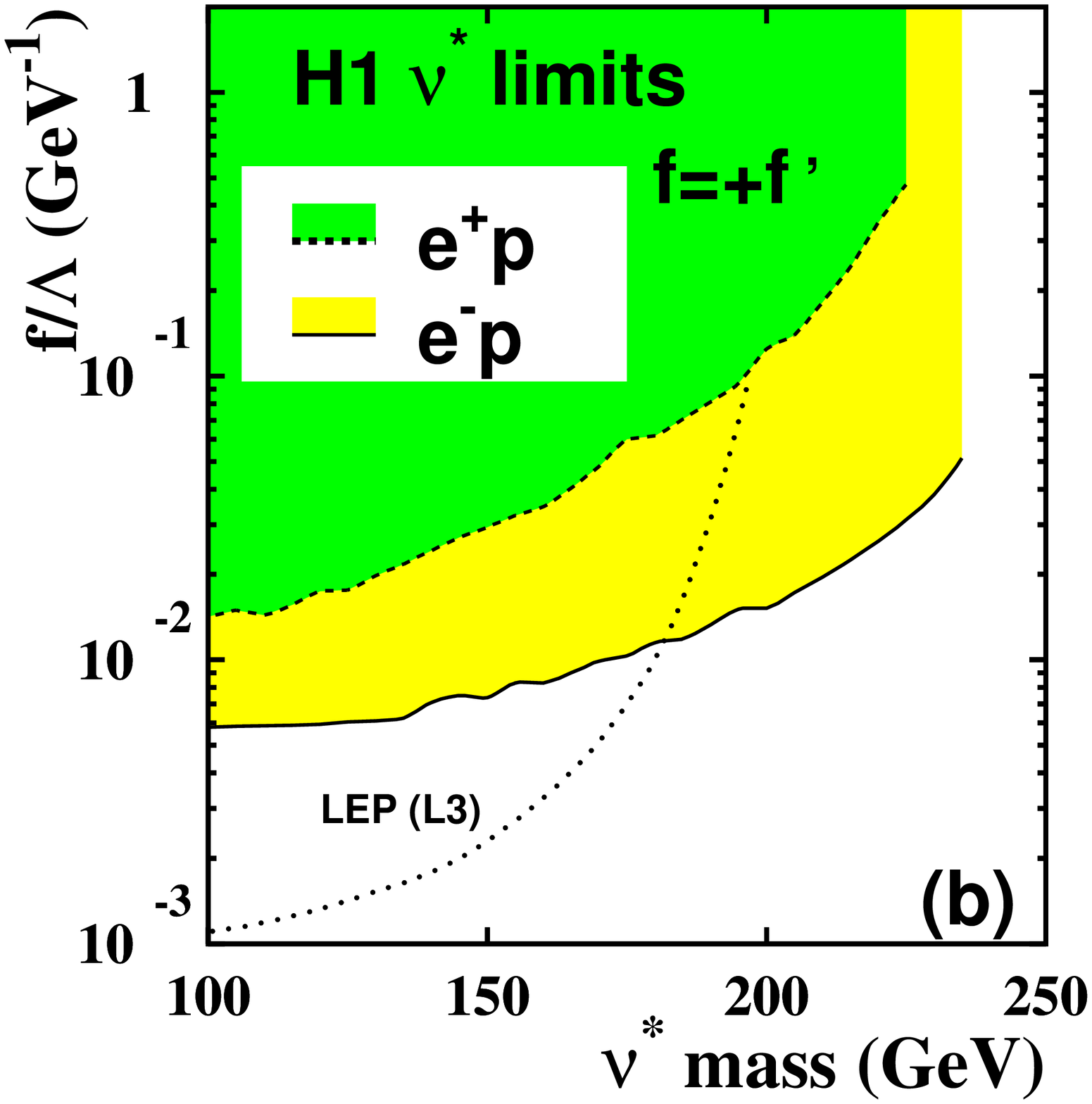} \\
\end{tabular}

\caption{Exclusion limits on the coupling $f/\Lambda$ at $95 \%$ confidence level
as a function of the mass
of excited neutrinos with the assumptions (a) $f=-f'$
and (b) $f=+f'$ .
Exclusion limits are given for H1 $e^-p$ data (full line) with an integrated luminosity of 15 pb$^{-1}$, for H1 $e^+p$ data~\cite{Adloff:2000gv} (dashed line) with an integrated luminosity of 37 pb$^{-1}$ and for L3~\cite{Acciarri:2001kb} (dotted line).}
\label{fig:folH1fm}

\end{center}
\end{figure}
The number of observed and expected events is counted within a sliding mass window which is adopted to the width of the expected excited neutrino signal.
Systematic uncertainties are taken into account as in~\cite{Adloff:2000gv}.

The resulting limits after combination of all decay channels
are given as a function of the $\nu^*$ mass in Fig.~\ref{fig:folH1fm}, for the
conventional assumptions $f=-f'$ and $f=+f'$. Note that
the decay \nnga is forbidden for $f=+f'$.
These results improve significantly our limits published earlier in $e^-p$~\cite{Ahmed:1994yi}
and $e^+p$~\cite{Adloff:2000gv} collisions and reach masses up to 
240 GeV and couplings $f/\Lambda$ of order $O(1/100 GeV)$. 
Using the
assumption $f/\Lambda = \frac{1}{M_{\nu^*}}$ excited neutrinos with masses between 50 GeV and
150 GeV (100 GeV and 140 GeV) are excluded by the H1 analysis for 
$f=-f'$ ($f=+f'$).

%Similar results were presented recently by ZEUS \cite{Chekanov:2001xk}, where
%however the narrow width approximation was used also at the highest masses.
Fig.~\ref{fig:folH1fm} also shows for comparison results obtained by
the L3 collaboration in $e^+e^-$ collisions at centre of mass energies
up to 202 GeV at LEP II~\cite{Acciarri:2001kb}.
The H1 limits are more stringent at high masses beyond the kinematic
reach of LEP II.  

\begin{figure}[tb]
\begin{center}
 \begin{tabular}{p{0.4\textwidth}p{1.0\textwidth}}
% -> ci-dessus : 0.3 = taille horizontale pour la
%    caption (a gauche). 0.7 = taille pour la fig
    \vspace*{1.5cm}
% -> modifier le vspace ci-dessous pour monter/descendre
%    la caption
      \caption[]{ \label{fig:folScan}
{Exclusion limits for excited neutrinos on the coupling $f/\Lambda$ at $95 \%$ confidence level
as a function of the value of $f'/f$. Each curve corresponds to a different $\nu^*$ mass. The circles indicate the maximum (worst limit) of each curve.
The areas above the lines are excluded.}}      &
\vspace*{-1.5cm}
      \mbox{\epsfxsize=0.50\textwidth
       \epsffile{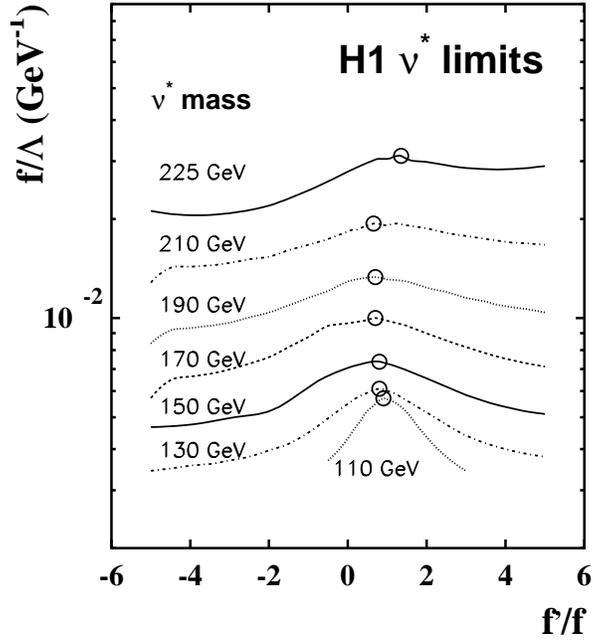}}
\end{tabular}
\end{center}
\end{figure}

\begin{figure}[htb]
\begin{center}
 \begin{tabular}{p{0.4\textwidth}p{1.0\textwidth}}
% -> ci-dessus : 0.3 = taille horizontale pour la
%    caption (a gauche). 0.7 = taille pour la fig
%    \vspace*{-7.5cm}
% -> modifier le vspace ci-dessous pour monter/descendre
%    la caption
      \caption[]{ \label{fig:folIndep}
{Exclusion limits on the coupling $f/\Lambda$ at $95 \%$ confidence level
as a function of the mass
of excited neutrinos. All $f'/f$ values in the interval 
$[-5;+5]$  have been considered (see figure~\ref{fig:folScan}), so this limit is independent of the relation between $f$ and $f'$ in that interval. 
The area above the line is excluded.}}      &
\vspace*{-1.5cm}
      \mbox{\epsfxsize=0.50\textwidth
       \epsffile{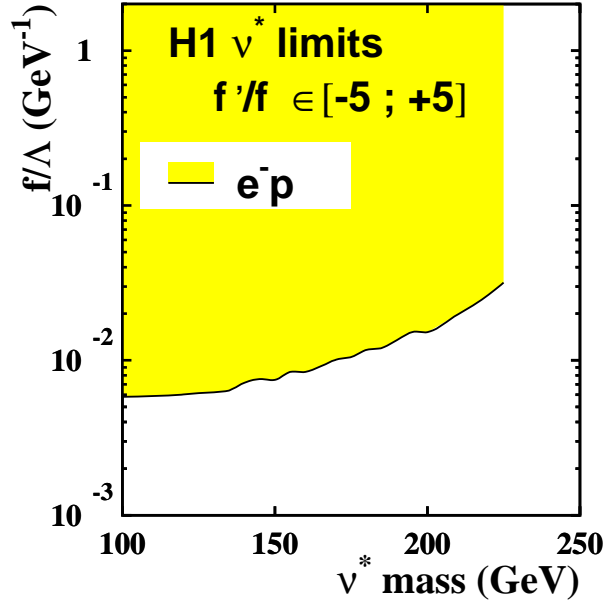}}
\end{tabular}
\end{center}
\end{figure}

Less model-dependent limits can be derived if arbitrary ratios $f'/f$
are considered. Fig.~\ref{fig:folScan} illustrates how the limits depend on this ratio
for various $\nu^*$ mass hypothesis. By choosing the point with the worst limit for each mass hypothesis, 
limits have been derived which are no longer dependent on $f'/f$ in
the range -5$< f'/f <$5. 
The result is shown in Fig.~\ref{fig:folIndep}. It deviates
from the limits obtained assuming $f=+f'$ only for high $\nu^*$ masses.
Limits on single $\nu^*$ production independent of $f'/f$ also have been shown previously by the
OPAL collaboration~\cite{Abbiendi:2000sa}.

%------------------------------------------------------------------------------
%\section{ Summary} 
In summary, using $e^- p$ data
a search for the production of excited neutrinos has been performed and no indication of a signal was found.
New limits have been established as function of couplings 
and excited neutrino masses both for specific relations between 
the couplings ($f=f'$ and $f=-f'$) and independent of the ratio of $f$ and $f'$. In comparison to previous analyses the data
presented here restrict the existence of excited neutrinos for masses up to 240
 GeV and to much smaller couplings.

%------------------------------------------------------------------------------

%%%%%%%%%%%%%%%%%%%%%%%%%%%%%%%%%%%%%%%%%%%%%%%%%%%%%%%%%%%%
%\section*{Acknowledgements}

We are grateful to the HERA machine group whose outstanding
efforts have made and continue to make this experiment possible. 
We thank
the engineers and technicians for their work in constructing and now
maintaining the H1 detector, our funding agencies for 
financial support, the
DESY technical staff for continual assistance
and the DESY directorate for the
hospitality which they extend to the non DESY 
members of the collaboration.

\clearpage

%%%%%%%%%%%%%%%%%%%%%%%%%%%%%%%%%%%%%%%%%%%%%%%%%%%%%%%%%%%%

\end{document}